\documentclass[%
reprint,
superscriptaddress,
 amsmath,amssymb,
 aps,
]{revtex4-2}

\usepackage[dvipdfmx]{graphicx}
\usepackage{dcolumn}
\usepackage{bm}

\usepackage[dvipdfmx]{color}
\usepackage{amsmath,amssymb}
\usepackage{unit}

\begin{document}

\preprint{APS/123-QED}

\title{Koopman mode decomposition of oscillatory temperature field inside a room}%
\thanks{Corresponding author: 
Yoshihiko Susuki}%

\author{Naoto Hiramatsu}%
\altaffiliation[Presently, ]{Daikin Industries, Ltd.}
\affiliation{%
Department of Electrical and Information Systems, Osaka Prefecture University, 
1-1 Gaken-cho, Naka-ku, Sakai 599-8531 Japan
}%
\author{Yoshihiko Susuki}%
\email{susuki@eis.osakafu-ac.jp}
\affiliation{%
Department of Electrical and Information Systems, Osaka Prefecture University, 
1-1 Gaken-cho, Naka-ku, Sakai 599-8531 Japan\\
JST, PRESTO, 4-1-8 Honcho, Kawaguchi, Saitama 332-0012, Japan
}%
\author{Atsushi Ishigame}%
\affiliation{%
 Department of Electrical and Information Systems, Osaka Prefecture University, 
 1-1 Gaken-cho, Naka-ku, Sakai 599-8531 Japan
}%

%
%

\date{\today}

\begin{abstract}
Koopman mode decomposition (KMD) is a technique of nonlinear time-series analysis capable of decomposing data on complex spatio temporal dynamics into multiple modes oscillating with single frequencies, called the Koopman modes (KMs). 
We apply KMD to measurement data on oscillatory dynamics of a temperature field inside a room that is a complex phenomenon ubiquitous in our daily lives and has a clear technological motivation in energy-efficient air conditioning.  
To characterize not only the oscillatory field (scalar field) but also associated heat flux (vector field), we introduce the notion of a temperature gradient using the spatial gradient of a KM.  
By estimating the temperature gradient directly from data, we show that KMD is capable of extracting a distinct structure of the heat flux embedded in the oscillatory temperature field, relevant in terms of air conditioning.  
\end{abstract}

\maketitle


\section{Introduction}

In this paper, we address dynamics of a temperature field arising in a room.  
The thermal dynamics are ubiquitous in our daily lives and typical of complex phenomena emerging as an interaction of physical and engineered systems such as fluid flows, air conditioners, and humans \cite{Bulk}.  
Exploring the dynamics is of fundamental interest in far-from-equilibrium physics and of technological significance in energy-efficient air conditioning, especially by \emph{in situ} diagnosis of temperature-field dynamics. 
The dynamics are traditionally studied with the model-based approach, that is, numerical simulations of nonlinear differential equations of fluid flows and so on, which is referred to as computational fluid dynamics (CFD): see, e.g., Refs.\,\citep{CFD2,CFD3}. 
Because CFD is normally time-consuming, it is not applicable to the \emph{in situ} diagnosis of temperature-field dynamics.  
Thus, instead of the model-based approach, an alternative data-based approach to the in-situ diagnosis has been desirable and feasible due to the development of internet-of-things technology; see, e.g., Refs.\,\cite{Wen:2018,Kono:2018}. 

Recently, the second author's group proposed the use of Koopman mode decomposition (KMD) for the dynamical analysis of heat transfer inside atriums in buildings \cite{Kono:2017a,Kono:2020}. 
KMD is a new technique of nonlinear time-series analysis based on eigenvalues of the so-called Koopman operator, which is a linear infinite-dimensional operator defined for a nonlinear dynamical systems \cite{Mezic:2005,Rowley:2009,Mezic:2013}. 
The technique is capable of decomposing multichannel time series into modal oscillations, each of which contains a single frequency, characterized by the eigenvalues. 
KMD has the solid mathematical foundation on an operator theory of dynamical systems \cite{Mezic:2005}, a clear property of timescale separation of complex spatiotemporal dynamics, and has been used widely in modeling and analysis of thermal dynamics in buildings \cite{Eisenhower:2010,Georgescu:2012,Georgescu:2015,Masaki:2019,Ljubi:2019}.  

The purpose of this paper is to utilize the capability of data-driven KMD for characterizing oscillatory dynamics of a temperature field inside a room.  
Specifically, we estimate the spatial derivative of such an oscillatory temperature field, which we will term a temperature gradient, directly from measurement data without development of underlying dynamic models.  
The estimation of a temperature gradient is direct to the \emph{in situ} diagnosis of heat flux inside a room according to the well-known Fourier's law \citep{fourier}, which is related to the conservation law of the total internal energy of working air. 
The estimation method is based on the so-called phase averaging that is widely used in fluid research \cite{phase} and connected to the Koopman operator framework in Ref.\,\cite{Mezic:2005}. 
We apply the method to time-series data on temperature measured in a practically used laboratory space and then show that KMD is capable of extracting a distinct structure of the heat flux embedded in the oscillatory temperature field, which is relevant in terms of air conditioning.  

The current application of KMD is novel in terms of the previous studies 
\cite{Eisenhower:2010,Georgescu:2012,Georgescu:2015,Kono:2017a,Masaki:2019,Ljubi:2019,Kono:2020}. 
The authors of Ref.\,\cite{Eisenhower:2010} focus on long-term (24\,h period) thermal dynamics for a whole building system, for which a minimal component is a room. 
The same focus appears in the papers \cite{Georgescu:2012,Georgescu:2015,Ljubi:2019}. 
In Refs.\,\cite{Kono:2017a,Masaki:2019,Kono:2020} the authors focus on short-term dynamics (less than 6\,h period) in building systems, while not focusing on the fine-scale spatial structure inside a room, namely, the temperature field.  
In this paper, we conduct KMD of short-term, fine-scale dynamics of a temperature field.

This paper is organized as follows. 
In Section~\ref{sec:method} we introduce the methodology utilized in this paper to analyze the oscillatory dynamics of a temperature field directly from measurement data. 
In Section~\ref{sec:data} we introduce the data on a temperature field measured inside a practically used room.  
Section~\ref{sec:results} is the main result of this paper and devoted to the data-based analysis of a temperature field oscillating in time and associated structure of heat flux. 
We summarize and outline future work in Section~\ref{sec:outro}.

\section{Methodology}
\label{sec:method}

We introduce KMD as the key technique in this paper and then derive a simple method for estimating the spatial gradient of a temperature field directly from measurement data via KMD. 

\subsection{Koopman operator and Koopman mode decomposition}

Consider a continuous-time, finite-dimensional dynamical system described by the following ordinary differential equation:
\begin{equation}
\frac{d\mathbf{x}}{dt}
=\mathbf{F}(\mathbf{x}), \qquad \forall\mathbf{x}\in\mathbb{R}^n,
\label{eq:2.1}
\end{equation}
where $\mathbf{x}\in\mathbb{R}^n$ is the state of the system, and $\mathbf{F}:\,\mathbb{R}^n\to\mathbb{R}^n$ is the nonlinear vector field. 
Under a proper condition of $\mathbf{F}$, the so-called flow $\mathbf{S}^t:\,\mathbb{R}^n\to\mathbb{R}^n$, $t\geq 0$ is defined as one parameter semigroup based on trajectories of the system \eqref{eq:2.1}. 
Here, we introduce a function defined on the state space, $f:\,\mathbb{R}^n\to{\mathbb{C}}$, and we call it an \emph{observable} while writing a space of all observables as $\mathbb{K}$. 
Then, the linear operator $U^t:\,\mathbb{K}\to\mathbb{K}$, $t\geq 0$ mapping an observable $f$ to new one is defined through the composition as 
\begin{equation}
U^tf := f\circ\mathbf{S}^t, \qquad \forall{f}\in\mathbb{K}.
\label{eq:2.2}
\end{equation}
This $U^t$ is called the \emph{Koopman operator} and an infinite-dimensional operator that represents the time evolution of $f$ under the flow $\mathbf{S}^t$ ($t\geq 0$).  
It completely keeps the information on the original nonlinear system \eqref{eq:2.1} under a certain condition, precisely, $\mathbb{K}$ contains the components of $\mathbf{x}$; see, e.g., Ref.\,\cite{TheBook:2020}. 

For the current analysis, we introduce the notion of eigenvalues and eigenfunctions of the Koopman operator $U^t$. 
A \emph{Koopman eigenvalue} (KE) is a complex number $\mu$ such that there exists a nonzero $\phi\in\mathbb{K}$, called a \emph{Koopman eigenfunction}, such that $U^t\phi=\exp(\mu t)\phi$.  
If the cardinality of KEs is countably infinite (see Ref.\,\cite{TheBook:2020}), we use the integer subscript $j$ to represent a pair of KE and a Koopman eigenfunction:
\begin{equation}
U^t{\phi}_j=\exp({\mu_j t}){\phi}_j \qquad \forall{t}\geq 0.
\label{eq:2.3}
\end{equation}

Now, we are in a position to introduce the so-called Koopman Mode Decomposition (KMD).  
By keeping in mind that multivariate time-series data on temperature field are analyzed in this paper, a vector-valued observable is defined as $\mathbf{f}:=[f_1,\ldots,f_M]^\top$ ($f_{i}\in\mathbb{K}$), where $\top$ stands for the transpose operation of vectors. 
The positive integer $M$ will be connected to the number of measurement locations of a temperature field later. 
For this observable $\mathbf{f}$, the positive time evolution of $\mathbf{f}$ along the flow $\mathbf{S}^t$ starting at the initial state $\mathbf{x}(0)$, denoted by $\mathbf{y}(t)=\mathbf{f}(\mathbf{S}^t(\mathbf{x}(0)))$, is expressed with the Koopman operator $U^t$ as follows:
\begin{equation}
\mathbf{y}(t)=[(U^tf_1)(\mathbf{x}(0)),\ldots,(U^tf_M)(\mathbf{x}(0))]^\top, \forall t\geq 0.
\label{eq:2.4}
\end{equation}
Here, by assuming as in Ref.\,\cite{Rowley:2009} that each $f_i$ is expanded with the Koopman eigenfunctions $\phi_j$, \eqref{eq:2.4} is 
described as follows:
\begin{equation}
\mathbf{y}(t)=\sum_{j=1}^\infty\exp({\mu_jt})\phi_j(\mathbf{x}(0))\mathbf{V}_{j},\qquad\forall t\geq 0.
\label{eq:2.5}
\end{equation}
Here, $\mathbf{V}_{j}\in\mathbb{C}^{M}$ is a constant complex-valued vector for the spectral expansion of $\mathbf{f}$ and is called a 
Koopman mode (KM), which was originally coined in Ref.\,\cite{Rowley:2009} for discrete-time dynamical systems. 
Equation \eqref{eq:2.5} shows that the possibly complicated time evolution $\mathbf{y}(t)$ driven by the nonlinear system \eqref{eq:2.1} can be decomposed into an infinite number of KMs, each of which oscillates at a single frequency characterized by KE.  
This type of spectral expansion of signals is called KMD. 

Before moving the estimation idea in the next section, we suppose that the multivariate signal $\mathbf{y}(t)$ is derived by synchronously measuring the target dynamics evolving in a spatial domain $D$ with dimension $d\in\{1,2,3\}$, and we denote the $M$ distinct locations for the measurement by $\mathbf{r}_{1},\ldots,\mathbf{r}_{M}\in\mathbb{R}^{D}$. 
Regarding this, (\ref{eq:2.5}) can be rewritten as
\begin{align}
\mathbf{y}(t) &=
\left[
\begin{array}{c}
y(t; \mathbf{r}_1) \\ \vdots \\ y(t; \mathbf{r}_M)
\end{array}
\right] \nonumber\\
&= \sum_{j=1}^\infty
\exp({\mu_jt})
\phi_j(\mathbf{x}(0))
\left[
\begin{array}{c}
V_j(\mathbf{r}_1) \\ \vdots\\ V_j(\mathbf{r}_M)
\end{array}
\right], \quad \forall t\geq 0.
\label{eq:2.6}
\end{align}
This expansion implies that the KM $\mathbf{V}_j$ is dependent on the location $\mathbf{r}$, and that the dynamics measured in space can be represented by the finite-dimensional system \eqref{eq:2.1}.  
This is mathematically tractable for several situations, i.e., as in Ref.\,\cite{Mezic:2005}, where the dynamics are sufficiently approaching to a finite-dimensional attractor, called an inertial manifold \cite{Temam:1997}, or where the dynamics can be approximately captured with a spatial discretization of an original infinite-dimensional dynamical system, the latter of which corresponds to our working assumption in this paper.

\subsection{Estimation of Koopman modes}

Estimation of KMs without knowledge of the nonlinear system \eqref{eq:2.1} is central to the data-based analysis in this paper. 
This type of estimation algorithm is generally referred to as dynamic mode decomposition \cite{Kutz:2016}. 
In this paper, to clarify both the physical and signal-processing perspectives of KM, we introduce the so-called phase averaging in Ref.\,\cite{phase} based on the signal-oriented representation \eqref{eq:2.6}.  
The connection between the phase averaging and KM is mentioned in Ref.\,\cite{Mezic:2005}. 
We now suppose that an oscillatory component of period $T(>0)$ is contained in the scalar-valued signal $y(t; \mathbf{r}_i)$ in \eqref{eq:2.6}.  
This implies that one KE, denoted by $\mu_\ell$, satisfies $\mu_\ell=\ii 2\pi/T$ ($\ii$ is the imaginary unit).  
Thus, the phase averaging $\langle{y(\mathbf{r}_i)}\rangle_T$ of $y(t; \mathbf{r}_i)$ with respect to period $T$ is defined as follows:
\begin{equation}
\langle{y(\mathbf{r}_i)}\rangle_{T}
:=\lim_{N \to \infty}\frac{1}{N}\sum_{k=0}^{N-1}y(kT; \mathbf{r}_i).
\label{eq:2.7}
\end{equation}
Here we assume that all the KEs are distinct. 
Because $\mathbf{y}(t)$ are derived by real-field measurement and thus are real-valued, there exists the KE that is the complex conjugate of $\mu_\ell$, denoted as $\mu_{\ell+1}=-\ii 2\pi/T$. 
Then, the phase averaging of the multivariate signal $\mathbf{y}(t)$ in \eqref{eq:2.6} with respect to $T$ is represented as follows:
\begin{equation}
\langle\mathbf{y}\rangle_T:=
\left[
\begin{array}{c}
\langle y(\mathbf{r}_1)\rangle_T \\ \vdots\\ \langle y(\mathbf{r}_M)\rangle_T
\end{array}
\right]
=\mathbf{\hat{V}}_\ell+\mathbf{\hat{V}}_{\ell+1},
\label{eq:2.8}
\end{equation}
with
\[
\mathbf{\hat{V}}_\ell
:=
\phi_\ell(\mathbf{x}(0))
\left[
\begin{array}{c}
V_\ell(\mathbf{r}_1) \\ \vdots\\ V_\ell(\mathbf{r}_M)
\end{array}
\right].
\]
Note that $\mathbf{\hat{V}}_\ell$ and $\mathbf{\hat{V}}_{\ell+1}$ are complex conjugate vectors, and their sum becomes real-valued. 
We will call the vector $\mathbf{\hat{V}}_\ell$ the KM, although it differs from the exact KM $\mathbf{V}_\ell$ possibly by a constant.  
Hence, provided that $\mu_\ell$, precisely $T$ is known with FFT etc., it is possible to estimate the sum of KMs, $\mathbf{\hat{V}}_\ell+\mathbf{\hat{V}}_{\ell+1}$, directly from the signal $\mathbf{y}(t)$ with the phase averaging. 

\subsection{Estimation of temperature gradient}
\label{subsec:main}

We are in a position to formulate the estimation problem of a temperature gradient directly from measurement data. 
For this, the multivariate signal $\mathbf{y}(t)$ in \eqref{eq:2.6} is regarded as time evolutions of a temperature field measured at the $M$ locations in a room.  
When denoting the original, continuum, temperature field by $\theta(t,\mathbf{r})$ ($t\geq 0$ and $\mathbf{r}\in D\subset\mathbb{R}^d$), we infer $\theta(t,\mathbf{r})$ from \eqref{eq:2.6} as 
\begin{equation}
\theta(t,\mathbf{r}) \sim\sum^\infty_{j=1}\exp(\mu_jt)\phi_j(\mathbf{x}(0))V_j(\mathbf{r})
\end{equation}
where $\phi_j(\mathbf{x}(0))$ is now a constant scalar. 
Then, by using the vector differential operator $\nabla_{\bf r}$ in Cartesian coordinates $\mathbf{r}$, the spatial derivative of the temperature field $\theta(t,\mathbf{r})$, namely, the temperature gradient, is derived as
\begin{equation}
\nabla_{\bf r}\theta(t,\mathbf{r}) \sim \sum^\infty_{j=1}\exp(\mu_jt)\phi_j(\mathbf{x}(0))\nabla_{\bf r}V_j(\mathbf{r}).
\end{equation}
This implies that in the KMD formulation, the location-dependent vector $\nabla_{\bf r}V_j(\mathbf{r})$ determines the temperature gradient with respect to the KE $\mu_j$, that is, the heat flux on the time scale determined by $\mu_j$.  

The location-dependent vector is related to the heat flux in the root-mean-square (RMS) sense. 
By defining the single mode component with the above KM with $\mu_\ell=\ii 2\pi/T$ embedded in $\theta(t,\mathbf{r})$ as
\begin{align}
\theta_{\{\ell,\ell+1\}}(t,\mathbf{r}) 
:=& \,\exp(\mu_\ell t)\phi_\ell(\mathbf{x}(0))V_\ell(\mathbf{r}) \nonumber\\
&+ \exp(\mu_{\ell +1}t)\phi_{\ell +1}(\mathbf{x}(0))V_{\ell +1}(\mathbf{r}) \nonumber\\
=& \,2\cos\left(\frac{2\pi}{T}t\right)\mathrm{Re}[\phi_\ell(\mathbf{x}(0))V_\ell(\mathbf{r})]  \nonumber\\
&- 2\sin\left(\frac{2\pi}{T}t\right)\mathrm{Im}[\phi_\ell(\mathbf{x}(0))V_\ell(\mathbf{r})],
\label{eqn:add}
\end{align}
we have the component-wise RMS vector of the temperature gradient $\nabla_{\bf r}\theta_{\{\ell,\ell+1\}}(t,\mathbf{r})$, denoted by $\langle\nabla_{\bf r}\theta_{\{\ell,\ell+1\}}(\mathbf{r})\rangle\sub{RMS}$ as follows: for each $\mathbf{r}=[r_1,\ldots,r_d]^\top\in D$,
\begin{align}
\langle\nabla_{\bf r}\theta_{\{\ell,\ell+1\}}(\mathbf{r})\rangle\sub{RMS} 
& \nonumber\\
&\makebox[-4em]{}:= \left[
\begin{array}{c}
\sqrt{\int^T_0\left\{{\DD\theta_{\{\ell,\ell+1\}}(t,\mathbf{r})}/{\DD{r_1}}\right\}^2{\dd t}/{T}} \\ \vdots \\
\sqrt{\int^T_0\left\{{\DD\theta_{\{\ell,\ell+1\}}(t,\mathbf{r})}/{\DD{r_d}}\right\}^2{\dd t}/{T}}
\end{array}
\right]
\nonumber\\
&\makebox[-4em]{}= \sqrt{2}|\phi_\ell(\mathbf{x}(0))|
\left[
\begin{array}{c}
|\DD V_\ell(\mathbf{r})/\DD{r_1}| \\ \vdots \\ |\DD V_\ell(\mathbf{r})/\DD{r_d}|
\end{array}
\right],
\end{align}
where $|\cdot|$ stands for the absolute value of numbers. 
This suggests that the weighted sum of location-dependent vectors, $\phi_\ell(\mathbf{x}(0))\nabla_{\bf r}V_\ell(\mathbf{r})+\phi_{\ell+1}(\mathbf{x}(0))\nabla_{\bf r}V_{\ell+1}(\mathbf{r})$, appears in the RMS-based estimation of the heat flux parameterized by the KE.

The estimation problem posed in this paper is to determine the weighed sum of location-dependent vectors. 
This can be solved with KMD in the above subsection. 
For computation, the multivariate signal $\mathbf{y}(t)$ in \eqref{eq:2.6} is treated as its sampled data in time, $\mathbf{y}_k$ ($k=0,1,\ldots$).  
The sampling period (in time) is assumed to be sufficiently smaller than the period $T$ of our interest. 
Given the KE $\mu_\ell$, equivalently $T$, the sum of KMs $\mathbf{\hat{V}}_\ell+\mathbf{\hat{V}}_{\ell+1}$  is computed directly from the time-series data with the phase averaging \eqref{eq:2.7}.  
From \eqref{eq:2.8}, the sum of KMs is regarded as the real-valued vector containing the values of the function $\phi_\ell(\mathbf{x}(0)){V}_\ell(\mathbf{r})+\phi_{\ell+1}(\mathbf{x}(0)){V}_{\ell+1}(\mathbf{r})$ at the $M$ locations $\mathbf{r}_1,\ldots,\mathbf{r}_M$.  
Therefore, in this paper we compute an approximation of the spatial derivative $\phi_\ell(\mathbf{x}(0))\nabla_{\bf r}{V}_\ell(\mathbf{r})+\phi_{\ell+1}(\mathbf{x}(0))\nabla_{\bf r}{V}_{\ell+1}(\mathbf{r})$ by applying the standard scheme of numerical differentiation (difference approximation) to the data $\mathbf{\hat{V}}_\ell+\mathbf{\hat{V}}_{\ell+1}$. 
The method of the estimation is summarized as the three steps after real-field measurement of temperature: (i) to determine the oscillation period $T$ embedded in the time-series data; (ii) to compute the sum of KMs with the phase averaging; and (iii) to compute the approximate spatial derivative.

\begin{figure}[t]
\centering
\includegraphics[width=0.45\textwidth]{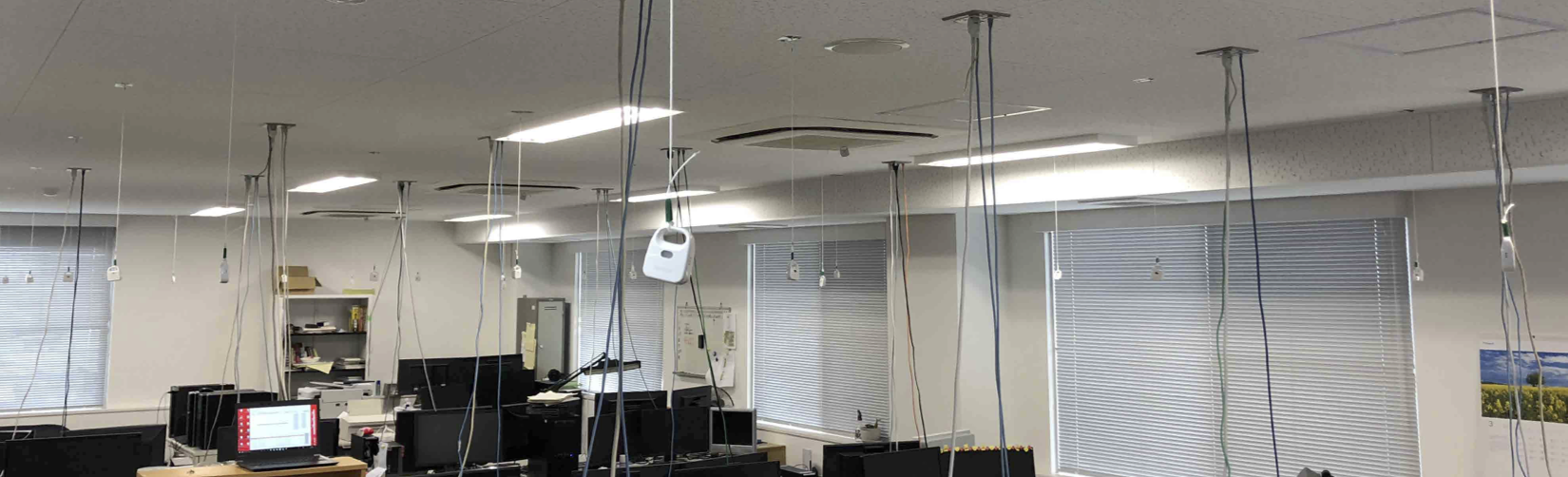} 
\vskip+5mm
\includegraphics[width=0.5\textwidth]{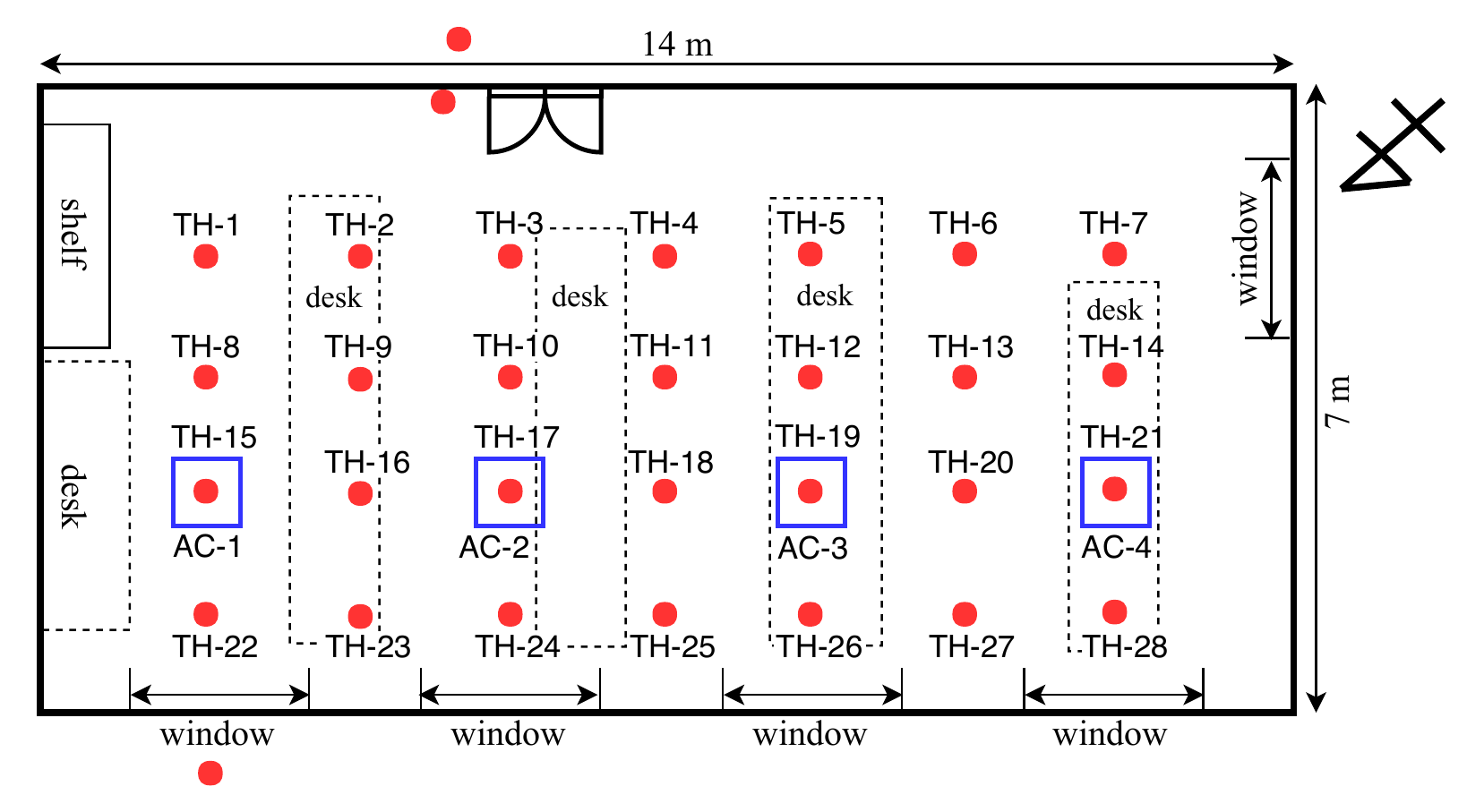} 
\caption{%
Photograph and geometry of room analyzed in this paper. 
The four \emph{blue} squares represent the air-conditioners equipped on the ceiling denoted by \textsf{AC-1} to \textsf{AC-4}. 
The \emph{red} dots denoted by \textsf{TH-1} to \textsf{TH-28} represent the environment sensors hung from the ceiling at a height of $1.9\U{m}$ from the floor, as shown in the photograph. 
The out-room temperature was measured at the two locations denoted by \emph{red} dots: one was close to the door, and the other was outside a window.
}%
\label{fig:3.1}
\end{figure}

\section{Measurement Data}
\label{sec:data}

In this section, we present time-series data on oscillatory dynamics of a temperature field measured inside a room. 
For this, we also introduce a practically used room for the measurement and its overview including the information on sensors.

\subsection{Target space}

The real-field measurement was conducted in the roomof our research group in Nakamozu Campus, Osaka Prefecture University, Sakai, Japan.  
Figure~\ref{fig:3.1} shows one photograph of the interior view of the space and its geometrical overview from the top of the space. 
The width of the room is about 14\,m, the depth is about 7\,m, and the height is about 2.6\,m. 
In this room, undergraduate and graduate students do research with desktop-type personal computers (PCs), where students and PCs are the main heat sources. 
Also, the four air-conditioners denoted by squares (\emph{blue}) in Figure~\ref{fig:3.1} are found on the ceiling.  
These air conditioners are called \textsf{AC-1} to \textsf{AC-4}, and each air conditioner  supplies air to its neighborhood. 
The room temperature was measured with environment sensors (OMRON, 2JCIE-BL01) that were introduced at 28 locations hung from the ceiling (at height 1.9\,m from the floor). 
These sensors are called \textsf{TH-1} to \textsf{TH-28} in Figure~\ref{fig:3.1}, where they are denoted with dots (\emph{red}).  
The out-room temperature was also measured at the two locations denoted with dots (\emph{red}): one was close to the door, and the other was outside a window. 
The measurement was conducted with the environmental sensor and data logger (HIOKI, LR50 Series).  

\begin{figure*}[t]
\centering
\includegraphics[width=0.85\textwidth]{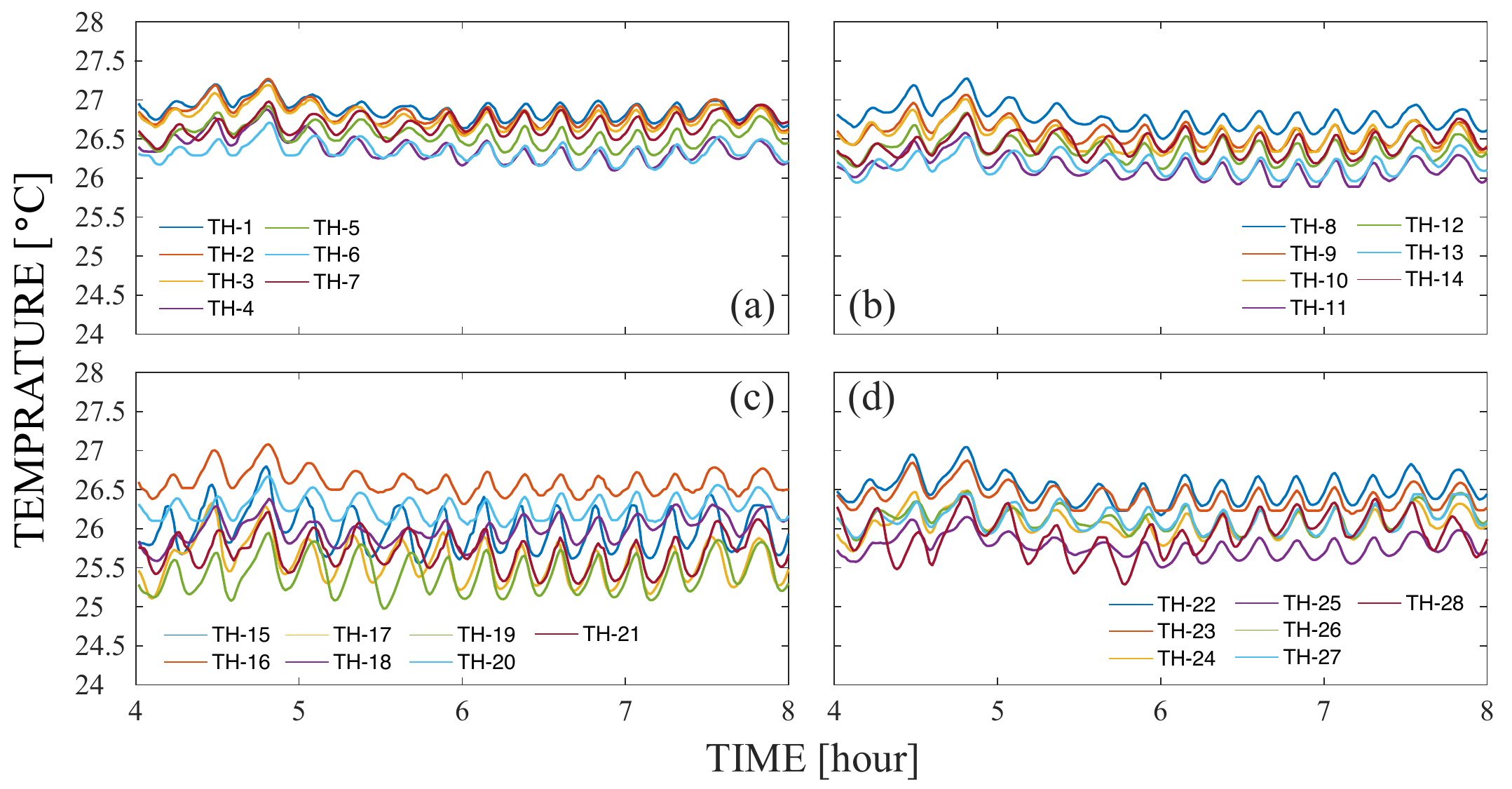} 
\caption{%
Time-series data on in-room temperature in summer. 
The data were measured between 4AM and 8AM (in Japan Standard Time) on July 17, 2018. 
The labels \textsf{TH-1} to \textsf{TH-28} correspond to the environment sensors shown in Figure~\ref{fig:3.1}. 
}%
\label{fig:3.2}
\end{figure*}

\begin{figure*}[t]
\centering
\includegraphics[width=0.85\textwidth]{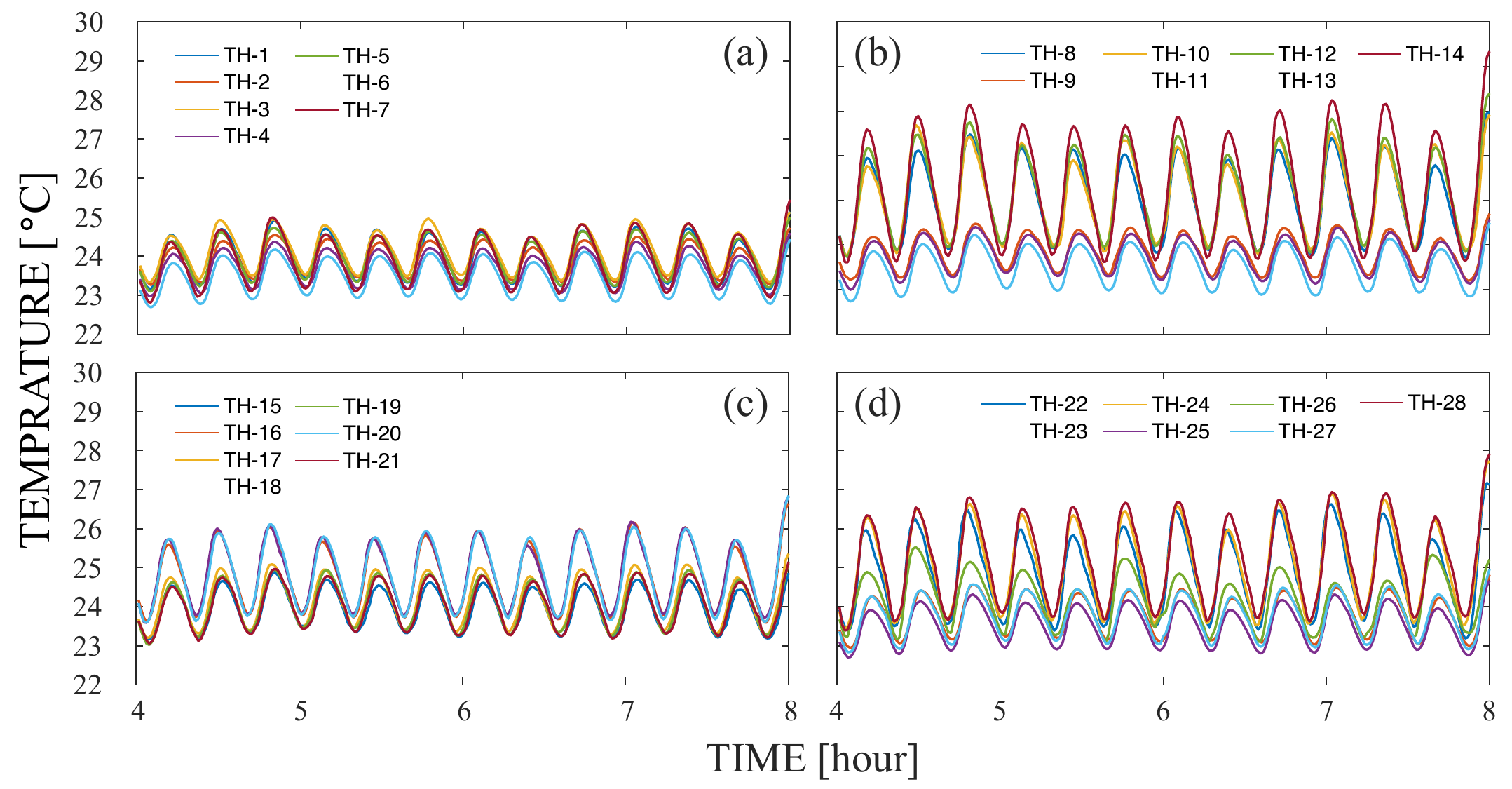} 
\caption{%
Time-series data on in-room temperature in winter. 
The data were measured between 4AM and 8AM (in Japan Standard Time) on January 9, 2019. 
The labels \textsf{TH-1} to \textsf{TH-28} correspond to the environment sensors shown in Figure~\ref{fig:3.1}. 
}%
\label{fig:3.3}
\end{figure*}

\subsection{Time-series data}

The measurement data are introduced for the two different seasons, summer and winter, in Japan. 
The air conditioners were operated during the measurement, and no one would work inside the room because the data shown below were collected in the early morning. 
This setting aims to simplify the temperature-field dynamics analyzed here by avoiding any disturbance due to the human presence. 
In this sense, the following phenomena shown in the data are relatively stationary and mainly affected by the intrinsic physical characteristics of the room and the operation of air conditioners.  
Also, the data for the different seasons will make it possible to estimate the season dependence of the underling heat flux. 

The data in summer are shown in Figure~\ref{fig:3.2} with a sampling period of 1\,min.  
In this figure, the total 28 colored lines represent the time evolutions of temperature at the 28 locations of the environment sensors, collected on Tuesday, July 17, 2018. 
The time duration of the data is from 4AM to 8AM in Japan Standard Time (JST). 
In this figure, we clearly see that a stationary oscillatory component is embedded in the time-series data. 
As mentioned above, the air-conditioners were operated continuously from 4AM to 8AM, and no one would work inside the room. 
It is thus anticipated that the oscillatory response in Figure~\ref{fig:3.2} results from the operation of air-conditioners with a feedback mechanism. 
The mechanism is basically that each air-conditioner turns on if it detects a too hot or cold temperature by sensors, which detailed information including control systems and parameters is not available for users. 

The data in winter are shown in Figure~\ref{fig:3.3} in the same manner as in Figure~\ref{fig:3.2}. 
The time duration of the data was again 4AM to 8AM (in JST) on Wednesday, January 9, 2019. 
By comparison with the summer data, the variation in temperature at several locations becomes larger in the winter data: for example, see panel (b) in Figures~\ref{fig:3.2} (summer) and \ref{fig:3.3} (winter). 
In Figure~\ref{fig:3.3}, we see again a stationary oscillatory component in the time-series data and anticipate the same cause as in summer---the oscillatory component is caused by an interaction between the physical room and the air conditioners.

\section{Results and Discussion}
\label{sec:results}

In this section, we apply KMD to the measured time-series data on temperature in Section~\ref{sec:data} and then estimate the temperature gradient in the real room using the method in Section~\ref{subsec:main}. 
Since the data at the 28 locations are sampled at the common height, the analysis of temperature field is regarded as a two-dimensional problem.  
This point will be mentioned later. 

\begin{table}[t]
\centering
\caption{%
Koopman mode decomposition of time-series data on in-room temperature in summer
}%
\vskip+2mm
\begin{tabular}{c c c c c} \hline
$\{j,j+1\}$ & $|\hat{\lambda}_{j}|$ & ${T}_{j} {\rm~[min.]}$ & $||\hat{\mathbf{V}}_{j}||$ & $E_j$ \\ \hline
\{1,2\} & 0.9913 & {\bf 14.23} & 1.0640 & 11.26\\
\{3,4\} & 0.9836 & 13.46 & 1.1740 & 9.356\\
\{5,6\} & 0.9738 & 89.16 & 1.6027 & 8.301\\
\{7,8\} & 0.9949 & 15.17 & 0.5542 & 7.461\\
\{9,10\} & 0.9176 & 7.182 & 1.6390 & 5.999\\
\{11,12\} & 0.9971 & 16.23 & 0.3205 & 5.193\\
\{13,14\} & 0.9924 & 18.63 & 0.4439 & 5.005\\
\{15,16\} & 1.0037 & 133.92 & 0.1349 & 4.921\\ \hline
\end{tabular}
\label{tab:4.1}
\end{table}

\begin{figure}[t]
\centering
\includegraphics[width=0.5\textwidth]{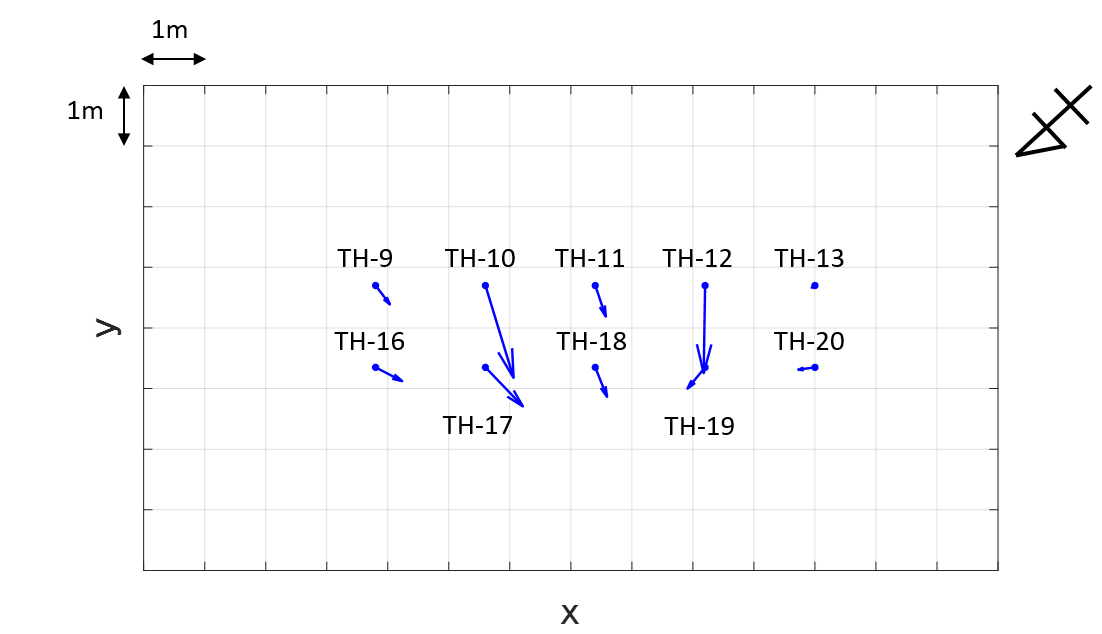} 
\caption{%
Visualization of temperature gradient in summer derived with KMD.  
The phase averaging with $T=14$\,min was used for the time-series data in Figure~\ref{fig:3.2}.
}%
\label{fig:4.1}
\end{figure}

\begin{table}[t]
\centering
\caption{%
Koopman mode decomposition of time-series data on in-room temperature in winter
}%
\vskip+2mm
\begin{tabular}{c c c c c} \hline
$\{j,j+1\}$ & $|\hat{\lambda}_{j}|$ & ${T}_{j} {\rm~[min.]}$ & $||\hat{\mathbf{V}}_{j}||$ & $E_j$ \\ \hline
\{1,2\} & 1.0060 & {\bf 18.99} & 0.9979 & 52.20\\
\{3,4\} & 0.9948 & 17.94 & 0.9395 & 12.44\\
\{5,6\} & 0.9288 & 2.912 & 2.4635 & 9.340\\
\{7,8\} & 0.8659 & 2.762 & 3.4384 & 9.332\\
\{9,10\} & 0.9890 & 13.09 & 0.9563 & 9.061\\
\{11,12\} & 1.0012 & 9.569 & 0.3457 & 8.797\\
\{13,14\} & 1.0042 & 20.63 & 0.2255 & 8.730\\
\{15,16\} & 0.9497 & 64.67 & 2.0646 & 7.841\\ \hline
\end{tabular}
\label{tab:4.2}
\end{table}

\begin{figure}[t]
\centering
\includegraphics[width=0.47\textwidth]{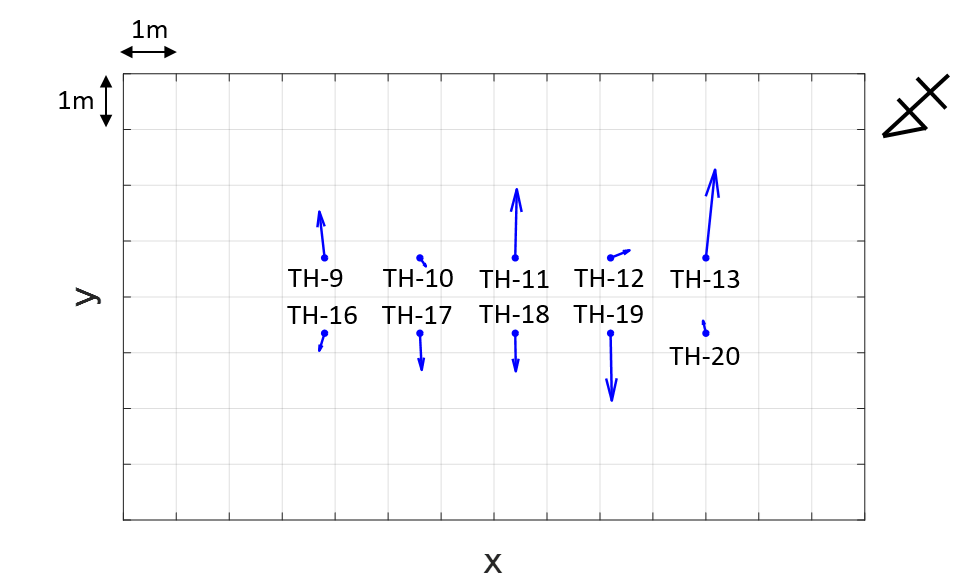} 
\caption{%
Visualization of temperature gradient in winter derived with KMD. 
The phase averaging with $T=19$\,min was used for the time-series data in Figure~\ref{fig:3.3}.
}%
\label{fig:4.4}
\end{figure}

\subsection{Summer data}

First, we apply KMD to the summer data in Figure~\ref{fig:3.2}. 
This aims to conduct the first step (i) in Section~\ref{subsec:main} for determining the oscillation period $T$ of the estimation.  
For this, we use the Arnordi-type algorithm \cite{Rowley:2009} to derive approximations of the KEs for the multivariate data $\mathbf{y}_k\in\mathbb{R}^M$ ($k=0,1,\ldots,N-1$) in Figure~\ref{fig:3.2}. 
The dimension $M$ of the data is equal to 28, and the number $N$ of snapshots from 4AM to 8AM is equal to 241. 
The Arnoldi-type algorithm derives the total 240 ($=N-1$) pairs of approximations of KEs $\hat{\lambda}_j:=\exp(\mu_j \times 60\,\U{s})$ and KMs ($\mathbf{\hat{V}}_j$).  
It is noted that the KEs $\hat{\lambda}_j$ are derived through the sampling (with period of 60\,s) of a continuous-time formulation  \eqref{eq:2.5} of KMD.  
To pick up a KE whose oscillation is stationary in time and dominant in the data, we introduce the energy-oriented norm $E_j$ from Refs.\,\cite{Kono:2017a,Kono:2020} as 
\begin{equation}
E_j := 
\begin{cases}
\sqrt{\displaystyle\sum_{k=0}^{N-1}||\hat{\lambda}^k_j\hat{\mathbf{V}}_j||^2} & (\hat{\lambda}_j\in\mathbb{R}),
\\\noalign{\vskip 2mm}
\sqrt{\displaystyle\sum_{k=0}^{N-1}||2{\rm Re}[\hat{\lambda}^k_j\hat{\mathbf{V}}_j]||^2} & ({\rm otherwise})
\end{cases}
\end{equation}
where $\|\cdot\|$ stands for the norm of vectors. 
Table \ref{tab:4.1} shows the derived KEs and KMs with high magnitude of the energy-oriented norm (except for the bias component in the original data).
The conjugate pairs of KEs are labeled as $\{j,j+1\}$ in this table.  
That is, the mode pair $\{1,2\}$ has the largest magnitude of the energy-orirented norm and thus the largest contribution to the data, which is regarded as a dominant mode. 
The oscillation period is about 14\,min, which is exactly confirmed from the summer data in Figure~\ref{fig:3.2}. 
Also, the absolute value of KE, $|\hat{\lambda}_{1,2}|$, is close to unity, and hence the corresponding oscillation is almost stationary. 

As the second and third steps (ii,iii) in Section~\ref{subsec:main}, we apply the phase averaging \eqref{eq:2.8} with $T=14\,\mathrm{min}$ to the multivariate data $\mathbf{y}_k$ in Figure~\ref{fig:3.2} and then derive an approximation of the spatial derivative $\phi_\ell(\mathbf{x}(0))\nabla_{\bf r}{V}_\ell(\mathbf{r})+\phi_{\ell+1}(\mathbf{x}(0))\nabla_{\bf r}{V}_{\ell+1}(\mathbf{r})$. 
The result of the spatial derivative is shown in Figure~\ref{fig:4.1}, where the gradient vectors at \textsf{TH-9} to \textsf{TH-13} and \textsf{TH-16} to \textsf{TH-20} are visualized.
As shown in Section~\ref{subsec:main}, the spatial derivative is related to the heat flux on a  timescale of 14\,min. 
According to Fourier's law \cite{fourier}, the gradient vectors should be assigned in a consistent manner that heat is transferred from one location (of air inside a room) with high temperature to another with low temperature. 
The original data in Figure~\ref{fig:3.2} were collected in Japan's summer, and thus the air conditioners \textsf{AC-2} and \textsf{AC-3} located at \textsf{TH-17} and \textsf{TH-19} were operated in the cooling mode. 
The cold air was supplied from the locations at \textsf{TH-17} and \textsf{TH-19}, and would affect the locations at \textsf{TH-10} and \textsf{TH-12}. 
The effect is clearly visualized with the gradient vectors derived here: see the \emph{downward} arrows at \textsf{TH-10} and \textsf{TH-12}. 
The downward direction implies that heat is transferred from the location at \textsf{TH-10} (or \textsf{TH-12}) to at \textsf{TH-17} (or \textsf{TH-19}), which is consistent with the cooling operation of the air conditioners.   

\subsection{Winter data}

Next, we address the winter data in Figure~\ref{fig:3.3} and present their KMD result in the same manner as above.  
The setting of $M$ and $N$ as well as the algorithm is the same as in the summer case. 
The derived KEs and KMs with a high magnitude of an energy-oriented norm are shown in Table~\ref{tab:4.2}. 
The mode pair $\{1,2\}$ with period of about 19\,min is regarded as the dominant mode.  
Also, the absolute value of KE, $|\hat{\lambda}_{1,2}|$, is close to unity, and hence the corresponding oscillation is almost stationary. 

The associated result on the spatial derivative is shown in Figure~\ref{fig:4.4}, where the gradient vectors at  \textsf{TH-9} to \textsf{TH-13} and \textsf{TH-16} to \textsf{TH-20} are again visualized. 
For the winter data in Figure~\ref{fig:3.3}, the air conditioners \textsf{AC-2} and \textsf{AC-3} located at \textsf{TH-17} and \textsf{TH-19} were operated in the heating mode. 
The hot air was supplied from the locations at \textsf{TH-17} and \textsf{TH-19}. 
The hot air tends to move along the ceiling in winter and directly affect the neighborhoods of the air-conditioners \textsf{AC-2} and \textsf{AC-3}. 
This implies that heat is transferred from the locations at \textsf{TH-17} and \textsf{TH-19} to their neighborhoods including the locations at \textsf{TH-9}, \textsf{TH-11}, and \textsf{TH-13}. 
In the measurement, all the sensors were hung close to the ceiling at a common height, and thus the measurement data are expected to capture the two-dimensional transfer of the hot air. 
This observation is consistent with the temperature gradient in Figure~\ref{fig:4.4}, where the \emph{upward} vectors appear at \textsf{TH-9}, \textsf{TH-11}, and \textsf{TH-13}. 

From the analyses, we conclude that KMD estimates the physically relevant structure of heat flux directly from the measurement data on temperature.

\section{Conclusion}
\label{sec:outro}

This paper is devoted to the data-based analysis of oscillatory dynamics of a temperature field inside a practically used room. 
The main technique is KMD applied to time-series data on temperature measured at multiple locations of sensors in the room. 
To characterize not only the oscillatory field (scalar field) but also associated heat flux (vector field), we introduced the notion of a temperature gradient using the spatial gradient of a KM.  
By estimating the gradient directly from data, we showed that KMD is capable of extracting a distinct structure of the heat flux embedded in the oscillatory temperature field, relevant in terms of air-conditioning. 
The method for estimating the gradient is basically to take the phase-average for time-series data and thus this can be simply implemented for the \emph{in situ} diagnosis of temperature-field dynamics. 

This study, which builds upon on the capability of KMD on thermal analysis in buildings demonstrated in early work \cite{Eisenhower:2010,Georgescu:2012,Georgescu:2015,Kono:2017a,Kono:2020}, is another step toward real-time control of temperature field for heat, ventilation, and air conditioning (HVAC) systems. 
In future work, based on the data-based analysis, we will synthesize a feedback controller for regulating the temperature field as a new function of HVAC. 
Concretely, using the estimated temperature gradient and KM, an output-feedback controller will be synthesized that is all data-driven without development of model. 
In this sense, this study will be a basis of the future work.

\vspace*{10mm}
\begin{acknowledgments}

We are grateful to Mr. Alexandre Ratieuville-Ogier (Osaka Prefecture University) for his reading of the paper. 
We also appreciate the referees for giving us suggestive comments. 
The work is supported in part by JST, CREST Grant No. JP-MJCR15K3 and JST, PRESTO Grant No. JP-MJPR1926. 

\end{acknowledgments}

%


\bibliography{reference}

\end{document}